\documentclass[12pt,a4paper]{article}
\usepackage{amsfonts}
\usepackage{mathrsfs}  
\usepackage{epsfig}
\usepackage{axodraw}
\usepackage{graphicx}

\def\simlt{\stackrel{<}{{}_\sim}}

\def\lrvec#1{\vbox{\ialign{##\crcr
${\hspace{1pt}\scriptscriptstyle\leftrightarrow\hspace{-1pt}}
$\crcr\noalign{\nointerlineskip}
$\hfil\displaystyle{#1}\hfil$\crcr}}}

\begin{document}

\title{Third order corrections to the ground state energy of the
  polarized diluted gas of spin $1/2$ fermions}
\author{\em  Piotr H. Chankowski, Jacek Wojtkiewicz and Rashad
  Bakhshizada\footnote{Emails:
    chank@fuw.edu.pl, wjacek@fuw.edu.pl, r.bakhshizad@student.uw.edu.pl}\\
Faculty of Physics, University of Warsaw,\\
Pasteura 5, 02-093 Warszawa, Poland
}
\maketitle
\abstract{We present the results of the computation of the
  third order corrections to the ground state energy of the diluted
  polarized gas of nonrelativistic spin $1/2$ fermions interacting
  through a spin-independent repulsive two-body potential.
  The corrections are computed within the effective field theory
  approach which does not require specifying the interaction
  potential explicitly but only to characterize it by only a few
  parameters - the scattering 
  lengths $a_0$, $a_1,\dots$ and effective radii $r_0,\dots$ -
  measurable in low energy fermion-fermion  elastic scattering.
  The corrections are computed semi-analytically, that is
  are expressed in terms of two functions of the system's polarization.
  The functions are given by the integrals which can be easily
  evaluated using the Mathematica built-in routines for numerical
  integration.
\vskip0.1cm

}

\newpage


\noindent{\bf Introduction.}
The classic model of the many-body quantum mechanics, the diluted gas of
nonrelativistic fermions interacting through a spin independent repulsive
two-body potential \cite{Kesio,FetWal} has attracted in the recent time a
renewed attention due to the advent of a new generation of experiments
with cold atomic gases in which the interaction strength can be tuned in
a wide range by exploiting the existence and properties of the so-called
Feshbach resonance \cite{CHGRJUTIE}. The experiments have stimulated
intensive numerical studies of the system \cite{QMC10,FRAPIL,QMC14} aiming
at computing its properties mainly relating to the possible application of
the model to the problem of the emergence of the so-called itinerant
ferromagnetism in systems of interacting fermions.

On the other hand, the application to the model of the effective field
theory method in the pioneering work \cite{HamFur00} (see also
\cite{FUHATI,HamFur02,HAKOL,KolckiSka}) has
opened new possibilities to investigate properties of the system of
interacting fermions analytically. The proposed approach has in particular
greatly simplified perturbative computations of the
ground-state energy of the system, automatically yielding its expansion
in powers of $k_{\rm F}R$ where $R$ is the lengths scale characterizing
the interaction potential and $k_{\rm F}$ is the Fermi wave vector of the
system of $N$ fermions enclosed in the volume $V$. 

The simplifications offered by the effective field theory approach allowed
to complete recently \cite{WeDrSch} the computation of the fourth order,
$(k_{\rm F}R)^4$, contribution to the ground-state energy of the
system of spin $s$ fermions with equal densities of fermions of different
spin projections (unpolarized) system. It also allowed to reproduce \cite{CHWO1}
semi-analytically but in the universal setting, that is without specifying
the underlying interaction potential, the order $(k_{\rm F}R)^2$
correction to the ground-state energy of the spin $1/2$ fermions
with the arbitrary ratio of the densities of spin up and spin
down fermions (arbitrarily polarized system) which in the past has been
computed by Kanno \cite{KANNO} using the hard spheres model interaction.
This result has recently been generalized to the system of spin $s$
fermions with arbitrary proportions of densities of the $g_s=2s+1$
possible spin projections \cite{PECABO}.

Computations of the ground-state energy as a function of the system's
polarization $P$ directly relates to the possibility of the phase
transition to the ordered state ($P\neq0$) at zero temperature with
increasing the strength of the interaction potential (reflected in the
effective field theory approach by the increasing magnitude of the
scattering lengths $a_0$, $a_1,\dots$ and the effective radii $r_0,\dots$)
and/or of the system's overall density characterized by its Fermi wave
vector $k_{\rm F}=(3\pi^2N/V)^{1/3}$. The first order of the perturbative
expansion (equivalent to the mean-field approximation) predicts that
such a transition occurs for $k_{\rm F}a_0=\pi/2$ (the Stoner's criterion
\cite{Stoner,Kesio}). Numerical investigations \cite{QMC10} which
necessarily use a concrete form of the interaction potential indicate
that the transition occurs at $k_{\rm F}a_0\approx0.85$. Inclusion of
the second order contribution in the perturbative expansion
of the ground state energy yields $k_{\rm F}a_0\approx1.054$ as the
critical value of the expansion parameter. 

In this letter we compute the third order correction to the ground-state
energy of the system of interacting spin $1/2$ fermions for an arbitrary
value of the system's polarization $P$. As in \cite{CHWO1} we apply the
effective field theory approach proposed first in \cite{HamFur00} and
regularize the divergent integrals over wave vectors with the help of the
cutoff $\Lambda$. We explicitly demonstrate the cancellation of the terms
diverging in the limit $\Lambda\rightarrow\infty$ after the couplings of
the effective theory Lagrangian are expressed in terms of the scattering
lengths computed up to the appropriate order using the same cutoff
$\Lambda$. The final result is expressed in terms of two functions of the
polarization $P$ which are given by the integrals which can be computed
with sufficient accuracy by the Mathematica package built-in routine for
numerical evaluation of the multidimensional integrals over a prescribed
domains.
\vskip0.1cm

\noindent{\bf Computation.}
Assuming that the underlying ``fundamental'' spin independent two-body
interaction of nonrelativistic spin $s$ fermions of mass $m_f$ is
consistent with the Galileo, parity and time-reversal symmetries, the
most general interaction term of the effective theory Hamiltonian which
captures properties of low density system of $N$ such fermions as well
as characteristics of their low energy scattering reads \cite{HamFur00}
\begin{eqnarray}
V_{\rm int}={C_0\over2}\!\int\!d^3\mathbf{x}
  \sum_{\alpha\beta}\psi_\alpha^\dagger\psi_\beta^\dagger\psi_\beta\psi_\alpha
  -{C_2\over16}\!\int\!d^3\mathbf{x}
  \sum_{\alpha,\beta}[\psi^\dagger_\alpha\psi^\dagger_\beta
    (\psi_\beta\lrvec{\mbox{\boldmath{$\nabla$}}}^2\!\psi_\alpha)+{\rm H.c.}]
 \phantom{aaaaa}~\nonumber\\
-~\!{C^\prime_2\over8}\!\int\!d^3\mathbf{x}\sum_{\alpha,\beta}
 (\psi^\dagger_\alpha\!\lrvec{\mbox{\boldmath{$\nabla$}}}\psi^\dagger_\beta)\!\cdot\!
  (\psi_\beta\!\lrvec{\mbox{\boldmath{$\nabla$}}}\psi_\alpha)
  +{D_0\over2}\!\int\!d^3\mathbf{x}\sum_{\alpha,\beta,\gamma}
  \psi_\alpha^\dagger\psi_\beta^\dagger\psi_\gamma^\dagger\psi_\gamma\psi_\beta\psi_\alpha
  +\dots\label{eqn:Heff}
\end{eqnarray}
$\psi_\alpha$ and $\psi^\dagger_\alpha$ are the usual field operators of the
second quantization formalism \cite{FetWal}. The coupling constants $C_0$,
$C_2,\dots$ multiplying the local operator structures of decreasing length
dimensions can be determined by computing using this interaction the
amplitude the elastic scattering of two fermions parametrized in the low
energy limit in terms of the scattering lengths. The result of such a
procedure is \cite{HamFur00,FUSTTI,CHWO1}
\begin{eqnarray}
  &&C_0={4\pi\hbar^2\over m_f}~\!a_0\left(1+{2\over\pi}~\!a_0\Lambda
  +{4\over\pi^2}~\!a^2_0\Lambda^2+\dots\right),\nonumber\\
  &&C_2={2\pi\hbar\over m_{\rm red}}~\!{1\over2}~\!a_0^2r_0+\dots,\phantom{aaaa}
  C_2^\prime={2\pi\hbar\over m_{\rm red}}~\!a_1^3+\dots\label{eqn:CouplingsDetermined}
\end{eqnarray}
$\Lambda$ is the UV cutoff imposed on the wave-vectors of the loop
integrals. Divergences, absent in the underlying ``fundamental''
theory, appear as a result of the  local (i.e. singular) nature of the
interaction terms of the effective interaction Hamiltonian
(\ref{eqn:Heff}).
\vskip0.1cm

The ground-state energy density of the system of $N$ noninteracting
nonrelativistic spin $s$ fermions (enclosed in the volume $V$) is
\begin{eqnarray}
  {E_{\Omega_0}\over V}={1\over6\pi^2}~\!\sum_{\alpha=1}^{g_s}{\hbar^2\over2m_f}~\!
  {3\over5}~\!p_{{\rm F}\alpha}^5~\!.\label{eqn:OthOrder}
\end{eqnarray}
$p_{{\rm F}\alpha}=((1/6\pi^2))N_\alpha/V)^{1/3}$ are the respective
Fermi wave vectors of $N_\alpha$ fermions with the spin projection $\alpha$
in the system; $N=\sum_{\alpha=1}^{g_s}N_\alpha$. Since energy of the system
of spin $1/2$ fermions is (in the absence of an external magnetic field)
invariant with respect to the interchange $N_\uparrow\leftrightarrow N_\downarrow$
we will in the following as in \cite{CHWO1}
denote $N_+$ (and, correspondingly, $p_{{\rm F}+}$)
the number of spin up fermions if $N_\uparrow\geq N_\downarrow$, and will
write the system's polarization $0\leq P\leq1$ as
\begin{eqnarray}
  P={N_+-N_-\over N_++N_-}\equiv{1-r^3\over1+r^3}~\!,
  \phantom{aaa}{\rm where}\phantom{aa}r\equiv{p_{{\rm F}-}\over p_{{\rm F}+}}
  ={N_-^{1/3}\over N_+^{1/3}}~\!.
\end{eqnarray}
It will be also convenient to express the results in terms of the
average Fermi wave-vector $k_{\rm F}=((6\pi^2/g_s)(N/V))^{1/3}$ which does
change when the numbers of fermions of different spin projections
are varied (keeping constant $N=N_++N_-$).

The first nontrivial correction to the ground-state energy has been
computed long time ago by Lenz \cite{Lenz}. Further corrections to
$E_\Omega$ are most easily computed using the general formula\footnote{The
  symbol T of the chronological ordering should not be confused with
  $T$ denoting time.} 
\begin{eqnarray}
  \lim_{T\rightarrow\infty}\exp(-iT(E_\Omega-E_{\Omega_0})/\hbar)=
  \lim_{T\rightarrow\infty}\langle\Omega_0
  |{\rm T}\exp\!\left(-{i\over\hbar}\!\int_{-T/2}^{T/2}\!dt~\!
  V^I_{\rm int}(t)\right)\!|\Omega_0\rangle~\!.\label{eqn:basicFormula}
\end{eqnarray}
in which $V^I_{\rm int}(t)$ is the interaction part of the theory Hamiltonian
taken in the interaction picture. In application to the considered system
this formula, which can be evaluated using the standard Dyson expansion,
gives the corrections $(E_\Omega-E_{\Omega_0})/V$ to the ground-state energy
density as a sum of the momentum space connected vacuum Feynman diagrams
(called in this context also the Hugenholtz diagrams) multiplied by $i\hbar$.

As the effective theory interaction (\ref{eqn:Heff}) consists of an
in principle infinite number of operator structures, diagrams which
should be taken into account to obtain the order $(k_{\rm F}R)^\nu$
contribution to $(E_\Omega-E_{\Omega_0})/V$ are selected by the
power counting rules \cite{HamFur00,Weinb}
\begin{eqnarray}
  \nu=5-\sum_iV_i\Delta_i=2+3L+\sum_iV_i(d_i-2)~\!,\label{eqn:PowerCountingRules}
\end{eqnarray}
in which $V_i$ is the number of the vertices of type $i$ with $d_i$
derivatives and $n_i$ lines attached to the vertex,
$L$ is the number of closed loops and $\Delta_i=5-d_i-{3\over2}n_i$
characterizes the dimension interaction vertices; $\Delta_{C_0}=-1$,
$\Delta_{C_1,C_2^\prime}=-3$, etc. Dimensional analysis
shows that the magnitude of the coupling $C_i$ multiplying the vertex
of type $i$ is $(4\pi\hbar^2/m_f)R^{-\Delta_i}$, where $R$ is the
characteristic length scale of the underlying interaction potential
(which, if in the assumed absence of any resonant or anomalous
behaviour, implies that all $a_\ell\sim r_\ell\sim R$). 
\vskip0.1cm

The power counting rules (\ref{eqn:PowerCountingRules}) tell that the
to the order $k_{\rm F}^5(k_{\rm F}R)^2$
correction to $(E_\Omega-E_{\Omega_0})/V$ contribute only diagrams with two
$C_0$ vertices and three loops. There is only one nonvanishing such diagram,
which has been first evaluated in \cite{HamFur00} for the case of unpolarized
system of spin $s$ fermions and shown to straightforwardly reproduce the
well-known classic result obtained first in \cite{HuangYang57} with the
help of rather cumbersome methods (and since then reproduced using a
variety of different approaches). In \cite{CHWO1} the corresponding three
loop diagram has been evaluated semi-analytically for the case of the
polarized system of spin 1/2 fermions and the result found to numerically
coincide with the analytic formula obtained by Kanno \cite{KANNO} within
the hard-spheres model of the two-body interaction (extension of the result
of \cite{CHWO1} to the arbitrarily polarized system of spin $s$ fermions
has been presented very recently in \cite{PECABO}).

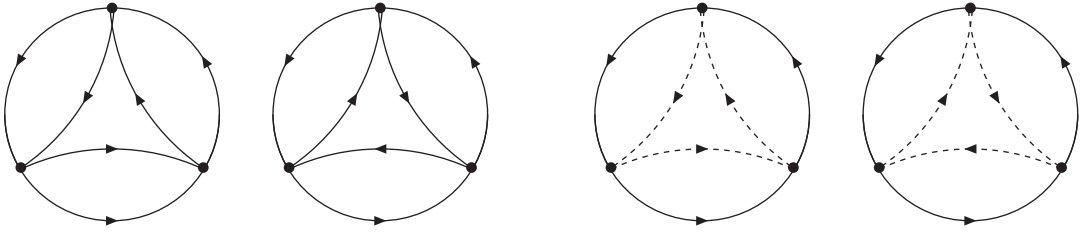
\begin{figure}[]
\begin{center}
\begin{picture}(400,100)(5,0)
  \ArrowArc(40,40)(40,330,90)
  \ArrowArc(40,40)(40,90,210)
  \ArrowArc(40,40)(40,180,360)
  \Vertex(74,20){2}
  \Vertex(6,20){2}
  \Vertex(40,80){2}
  \ArrowArcn(40,-53)(80,115,65)
  \ArrowArcn(119,86)(80,235,185)
  \ArrowArcn(-39,86)(80,355,305)
  \ArrowArc(140,40)(40,330,90)
  \ArrowArc(140,40)(40,90,210)
  \ArrowArc(140,40)(40,180,360)
  \Vertex(174,20){2}
  \Vertex(106,20){2}
  \Vertex(140,80){2}
  \ArrowArc(140,-53)(80,65,115)
  \ArrowArc(219,86)(80,185,235)
  \ArrowArc(61,86)(80,305,355)
  \ArrowArc(260,40)(40,330,90)
  \ArrowArc(260,40)(40,90,210)
  \ArrowArc(260,40)(40,180,360)
  \Vertex(294,20){2}
  \Vertex(226,20){2}
  \Vertex(260,80){2}
  \DashArrowArcn(260,-53)(80,115,65){2}
  \DashArrowArcn(339,86)(80,235,185){2}
  \DashArrowArcn(181,86)(80,355,305){2}
  \ArrowArc(360,40)(40,330,90)
  \ArrowArc(360,40)(40,90,210)
  \ArrowArc(360,40)(40,180,360)
  \Vertex(394,20){2}
  \Vertex(326,20){2}
  \Vertex(360,80){2}
  \DashArrowArc(360,-53)(80,65,115){2}
  \DashArrowArc(439,86)(80,185,235){2}
  \DashArrowArc(281,86)(80,305,355){2}
\end{picture}
\end{center}
\caption{The two nonvanishing order $C^3_0$ effective theory connected
  vacuum diagrams (the ``particle-particle'' diagram and the ``particle-hole''
  diagram) contributing to the correction $E^{(3)}_\Omega$ to the
  ground-state energy of the system of spin $s$ fermions with equal densities
  of different spin projections and their counterparts in the case
  of the system of spin $1/2$ fermions and $N_\uparrow\neq N_\downarrow$. In this
  second case solid and dashed lines represent propagators of fermions with
opposite spin projections.}
\label{fig:C0cubes}
\end{figure}

The order $k_{\rm F}^5(k_{\rm F}R)^3$ correction is given by the Hugenholtz
diagrams with either three $C_0$ vertices and four loops or by two-loop
diagrams with a single $C_2$ or $C_2^\prime$ vertex. There are only two
nonvanishing diagrams of the first kind \cite{HamFur00} shown in
Figure \ref{fig:C0cubes}. Both these diagrams come with the combinatoric
factor of 2 (when the interaction term of spin $1/2$ fermions is written as
$C_0\int\!\psi^\dagger_+\psi_+\psi^\dagger_-\psi_-$)
and both are given by an integral of a product of three identical
blocks which consist of two terms each; of the arising $2^3=8$ terms two
vanish as a result of the integration while the remaining six give rise
to only two different terms; the factor $2\cdot3$ cancels the factor $1/3!$
arising from the expansion of the exponent. 

After the standard steps (for more details see \cite{CHWO1})
the contribution of the ``particle-particle'' diagram to the energy
density can be written in the form
\begin{eqnarray}
  {E_\Omega^{(3){\rm p-p}}\over V}={128m_f^2C_0^3\over(2\pi)^8\hbar^4}
  \left[G^{(1)}(p_{{\rm F}-},p_{{\rm F}+})+G^{(2)}(p_{{\rm F}-},p_{{\rm F}+})\right],
  \label{eqn:C03ppEnergyPol}
\end{eqnarray}
with
\begin{eqnarray}
 G^{(1)}(p_{{\rm F}-},p_{{\rm F}+})=\int_0^{s_{\rm max}}\!ds~\!s^2~\!
  {1\over4\pi}\!\int\!d^3\mathbf{t}~\!
  \theta(p_{{\rm F}-}-|\mathbf{t}+\mathbf{s}|)
  \theta(p_{{\rm F}+}-|\mathbf{t}-\mathbf{s}|)
~\!(g(t,s))^2~\!,\label{eqn:G1FunctionDef}
\end{eqnarray}
\begin{eqnarray}
 G^{(2)}(p_{{\rm F}-},p_{{\rm F}+})=\int_0^{s_{\rm max}}\!ds~\!s^2~\!
  {1\over4\pi}\!\int\!d^3\mathbf{t}~\!
  \theta(|\mathbf{t}+\mathbf{s}|-p_{{\rm F}-})
  \theta(|\mathbf{t}-\mathbf{s}|-p_{{\rm F}+})
~\!(h(t,s))^2~\!.\label{eqn:G2FunctionDef}
\end{eqnarray}
where the functions $g(t,s)\equiv g(|\mathbf{t}|,|\mathbf{s}|)$ and
$h(t,s)\equiv h(|\mathbf{t}|,|\mathbf{s}|)$ are given below. The analogous
contribution of the ``particle-particle'' diagram to the energy density
of the unpolarized system of spin $s$ fermions is obtained by multiplying
(\ref{eqn:C03ppEnergyPol}) by the spin factor ${1\over2} g_s(g_s-1)$
and setting $p_{{\rm F}-}=p_{{\rm F}+}=k_{\rm F}$.

The function $g(t,s)$ is the one which appeared in \cite{CHWO1} in
evaluating the order $C_0^2$ contribution to the energy density; it can
be written in the form
\begin{eqnarray}
  g(t,s)=-\Lambda+g_{\rm fin}(t,s)+{t^2\over\Lambda}+
  {\cal O}(1/\Lambda^2),\label{eqn:g(t,s)General}
\end{eqnarray}
in which $\Lambda$ is the UV cutoff imposed on the divergent integral
over the wave vectors. The finite part $g_{\rm fin}(t,s)$ is 
for $0<s\leq{1\over2}(p_{{\rm F}+}-p_{{\rm F}-})$ given by
\begin{eqnarray}
  g(t,s)={1\over2}~\!p_{\rm F+}
  +{t\over4}\ln{(p_{\rm F+}-t)^2-s^2\over(p_{\rm F+}+t)^2-s^2}
  +{p^2_{\rm F+}-s^2-t^2\over8s}\ln{(p_{\rm F+}+s)^2-t^2\over(p_{\rm F+}-s)^2-t^2}~\!,
  \phantom{a}~\label{eqn:g(t,s)Explicit1}
\end{eqnarray}
and for ${1\over2}(p_{{\rm F}+}-p_{{\rm F}-})<s\leq s_{\rm max}$ by
\begin{eqnarray}
  g(t,s)={1\over4}(p_{\rm F+}+p_{\rm F-}+2s)
  +{t\over4}\ln{p_{\rm F+}+s-t\over p_{\rm F+}+s+t}
  +{t\over4}\ln{p_{\rm F-}+s-t\over p_{\rm F-}+s+t}\phantom{aaaaaa}~\nonumber\\
 +~\!{p^2_{\rm F+}-t^2-s^2\over8s}\ln{(p_{\rm F+}+s)^2-t^2\over u_0^2-t^2}
 +{p^2_{\rm F-}-t^2-s^2\over8s}\ln{(p_{\rm F-}+s)^2-t^2\over u_0^2-t^2}~\!,
 \phantom{a}~
 \label{eqn:g(t,s)Explicit2}
\end{eqnarray}
where
\begin{eqnarray}
  u_0^2={1\over2}(p_{{\rm F}+}^2+p_{{\rm F}-}^2)-s^2~\!.\label{eqn:u0squared}
\end{eqnarray}
Compared to the form of $g(t,s)$ given in \cite{CHWO1} we have retained
in (\ref{eqn:g(t,s)General}) the term proportional to $1/\Lambda$ to
show explicitly that the finite contributions of such terms (which are
absent in the dimensional regularization used in \cite{HamFur00})
cancel out. $h(t,s)$ is a new function given by the finite integral
\begin{eqnarray}
  h(t,s)={1\over4\pi}\!\int\!d^3\mathbf{u}~\!
  {\theta(p_{{\rm F}-}-|\mathbf{u}+\mathbf{s}|)
    \theta(p_{{\rm F}+}-|\mathbf{u}-\mathbf{s}|)\over
    \mathbf{t}^2-\mathbf{u}^2-i0}~\!.\label{eqn:h(t,s)functionDef}
\end{eqnarray}
Its analytic form
\begin{eqnarray}
  h(t,s)=-{1\over2}~\!p_{{\rm F}-}
  -{t\over4}\ln{t-(p_{{\rm F}-}-s)\over t+(p_{{\rm F}-}-s)}
  -{t\over4}\ln{t-(p_{{\rm F}-}+s)\over t+(p_{{\rm F}-}+s)}
  \phantom{aaaaaaaa}\nonumber\\
  +~\!{t^2-(p_{{\rm F}-}^2-s^2)\over8s}\ln{t^2-(p_{{\rm F}-}+s)^2\over
    t^2-(p_{{\rm F}-}-s)^2}~\!,\phantom{aaaaaaaaaaaaaaaaaa}
  \label{eqn:h(t,s)Explicit:slts0}
\end{eqnarray}
for $s<s_0$ and
\begin{eqnarray}
  h(t,s)={1\over2}\left(2s-p_{{\rm F}-}-p_{{\rm F}+}\right)
  -{t\over4}\ln{t-(p_{{\rm F}-}-s)\over t+(p_{{\rm F}-}-s)}
  -{t\over4}\ln{t-(p_{{\rm F}+}-s)\over t+(p_{{\rm F}+}-s)}
  \phantom{aaaaaa}\nonumber\\
    -{1\over8s}\left[(p_{{\rm F}+}-s)^2+(p_{{\rm F}-}-s)^2-2u_0^2\right]
    \phantom{aaaaaaaaaaaaaaaaaaaaaaaaaaa}\label{eqn:h(t,s)Explicit:s0gts}\\
    -~\!{t^2-p_{{\rm F}+}^2+s^2\over8s}
    \ln{t^2-(p_{{\rm F}+}-s)^2\over t^2-u_0^2}
   -{t^2-p_{{\rm F}-}^2+s^2\over8s}
   \ln{t^2-(p_{{\rm F}-}-s)^2\over t^2-u_0^2}~\!,\nonumber
\end{eqnarray}
for $s_0\leq s\leq s_{\rm max}$, can be obtained by the same technique, introduced
in \cite{FUSTTI}, which served to obtain the function $g(t,s)$.
Both these functions vanish for $s>s_{\rm max}={1\over2}(p_{{\rm F}+}+p_{{\rm F}-})$,
therefore the integrals over $s=|\mathbf{s}|$ in (\ref{eqn:G1FunctionDef})
and (\ref{eqn:G2FunctionDef}) are finite. Similarly manifestly finite
is the integral over $t=|\mathbf{t}|$ in (\ref{eqn:G1FunctionDef})
while the analogous integral in (\ref{eqn:G2FunctionDef})
is finite owing to the fact that $h(t,s)\sim1/t^2$ as $t\rightarrow\infty$.

The contribution to the energy density of the ``particle-hole'' diagram
of Figure \ref{fig:C0cubes} can be written in the form
\begin{eqnarray}
  {E_\Omega^{(3){\rm p-h}}\over V}=-{32m_f^2C_0^3\over(2\pi)^8\hbar^4}
  \left[K^{(1)}(p_{{\rm F}-},p_{{\rm F}+})+K^{(2)}(p_{{\rm F}-},p_{{\rm F}+})\right],
  \label{eqn:C03phEnergyPol}
\end{eqnarray}
(the corresponding contribution of the left ``particle-hole'' diagram
of Figure \ref{fig:C0cubes} to the energy density of the unpolarized
system of spin $s$ fermions is obtained by multiplying (\ref{eqn:C03phEnergyPol})
by the spin factor ${1\over2}g_s(g_s-1)(3-g_s)$ and setting
$p_{{\rm F}-}=p_{{\rm F}+}=k_{\rm F}$). The functions $K^{(1)}(p_{{\rm F}-},p_{{\rm F}+})$
and $K^{(2)}(p_{{\rm F}-},p_{{\rm F}+})$ are given by
\begin{eqnarray}
  K^{(1)}(p_{{\rm F}-},p_{{\rm F}+})=\int_0^\infty\!ds~\!s^2~\!{1\over4\pi}\!
  \int\!d^3\mathbf{t}~\!\theta(|\mathbf{t}+\mathbf{s}|-p_{{\rm F}-})
  \theta(p_{{\rm F}+}-|\mathbf{t}-\mathbf{s}|)~\!
  (f_1(\mathbf{t}\!\cdot\!\mathbf{s},~\!s))^2~\!,\phantom{a}
  \label{eqn:K1functionDef}
\end{eqnarray}
\begin{eqnarray}
  K^{(2)}(p_{{\rm F}-},p_{{\rm F}+})=\int_0^\infty\!ds~\!s^2~\!{1\over4\pi}\!
  \int\!d^3\mathbf{t}~\!\theta(p_{{\rm F}-}-|\mathbf{t}+\mathbf{s}|)
  \theta(|\mathbf{t}-\mathbf{s}|-p_{{\rm F}+})~\!
  (f_2(\mathbf{t}\!\cdot\!\mathbf{s},~\!s))^2~\!,\phantom{a}
  \label{eqn:K2functionDef}
\end{eqnarray}
and the functions $f_1(\mathbf{t}\!\cdot\!\mathbf{s},~\!s)$ and
$f_2(\mathbf{t}\!\cdot\!\mathbf{s},~\!s)$ are given by the integrals
\begin{eqnarray}
  f_1(\mathbf{t}\!\cdot\!\mathbf{s},~\!s)={1\over4\pi}\!\int\!d^3\mathbf{u}~\!
  {\theta(p_{{\rm F}-}-|\mathbf{u}+\mathbf{s}|)
    \theta(|\mathbf{u}-\mathbf{s}|+p_{{\rm F}+})\over(\mathbf{u}-\mathbf{t})\cdot
    \mathbf{s}+i0}~\!,\label{eqn:f1(t,s)Def}
\end{eqnarray}
and 
\begin{eqnarray}
  f_2(\mathbf{t}\!\cdot\!\mathbf{s},~\!s)={1\over4\pi}\!\int\!d^3\mathbf{u}~\!
  {\theta(|\mathbf{u}+\mathbf{s}|-p_{{\rm F}-})
    \theta(p_{{\rm F}+}+|\mathbf{u}-\mathbf{s}|)\over(\mathbf{u}-\mathbf{t})\cdot
    \mathbf{s}-i0}~\!.\label{eqn:f2(t,s)Def}
\end{eqnarray}
Both integrals defining the functions $f_1$ and $f_2$ are over
manifestly finite domains: the one defining $f_1$ is over the interior of the
ball of radius $p_{{\rm F}-}$ and exterior of the sphere of radius $p_{{\rm F}+}$
and the one defining $f_2$ - the other way around.
In $K_1(p_{{\rm F}-},p_{{\rm F}+})$ (\ref{eqn:K1functionDef}) the function $f_1$
is then integrated (over $d^3\mathbf{t}$) again over a manifestly
finite domain - namely over the interior of the ball of radius $p_{{\rm F}+}$
and the exterior of the sphere of radius $p_{{\rm F}-}$ while the function
$f_2$ is in $K_2(p_{{\rm F}-},p_{{\rm F}+})$ (\ref{eqn:K2functionDef})
integrated over the interior of the ball of $p_{{\rm F}-}$ and the
exterior of the sphere of radius $p_{{\rm F}+}$. The straightforward
analysis  shows that the poles at
$\mathbf{t}\cdot\mathbf{s}=\mathbf{u}\cdot\mathbf{s}$ are never
within the integration domains. Hence the factors $\pm i0$ are irrelevant.
It is also clear that $K_2(k_{\rm F},k_{\rm F})=K_1(k_{\rm F},k_{\rm F})$.

The most difficult part of the computation is obtaining analytical
expressions for the functions $f_1$ and $f_2$. The formulae for
$f_1$ (for $f_2$) have been obtained by shifting the center of the
$\mathbf{u}$-space in the regime of small $s$ to the center
of the $p_{{\rm F}+}$-sphere (of the $p_{{\rm F}-}$-sphere) and to the
center of the $p_{{\rm F}-}$-sphere (of the $p_{{\rm F}+}$-sphere)
in the regime of large $s$, introducing then the polar coordinated and
taking the resulting integrals analytically with the help of the
Mathematica routines; the results of the symbolic integrations
have been then simplified manually by exploiting the relations which follow
from the definitions of the integration domains (details will be
published elsewhere \cite{CHWO4}).
In this way we have arrived at
\begin{eqnarray}
  f_1(\mathbf{t}\!\cdot\!\mathbf{s},~\!s)
  ={1\over2s}\times\!\left\{\matrix{f_1^{(a)}(t~\!\eta-s,~\!s)\cr
    f_1^{(b)}(t~\!\eta+s,~\!s)\cr f_1^{(c)}(t~\!\eta+s,~\!s)}\right.~\!,
  \phantom{aaaa}
    f_2(\mathbf{t}\!\cdot\!\mathbf{s},~\!s)
  ={1\over2s}\times\!\left\{\matrix{f_2^{(a)}(t~\!\eta+s,~\!s)\cr
    f_2^{(b)}(t~\!\eta-s,~\!s)\cr f_2^{(c)}(t~\!\eta-s,~\!s)}\right.~\!,
  \label{eqn:f12functions}
\end{eqnarray}
where
\begin{eqnarray}
  f_1^{(a)}(t,s)=-2s^2+{t\over2}~\!(p_{{\rm F}+}-p_{{\rm F}-}-2s)-s~\!p_{{\rm F}-}
  -s~\!p_{{\rm F}+}~\!\xi_0
  \phantom{aaaaaaaaaaaaaaaaaaaaaaaaaaaaaa}\nonumber\\
  -{p_{{\rm F}+}^2\over2}
  \ln\!\left({t-p_{{\rm F}+}\xi_0\over t+p_{{\rm F}+}}\right)
  +{t^2\over2}\ln\!\left({t+2s+p_{{\rm F}-}\over t+p_{{\rm F}+}}\right)
  \phantom{aaaaaaaaaaaaaaaaaaaaaaaaaaaaaaa}~\nonumber\\
    +{1\over4}~\!(p_{{\rm F}-}^2-4s^2-4s t)\left\{
      -2\ln\!\left({p_{{\rm F}+}+4s\xi_0\over p_{{\rm F}-}-2s}\right)\right.
      \phantom{aaaaaaaaaaaaaaaaaaaaaaaaaaaaaaaa}~\nonumber\\
      \left.+\ln{[t^2-(p_{{\rm F}-}^2-4s^2-4st)\xi_0^2]
       [tp_{{\rm F}+}-(p_{{\rm F}-}^2-4s^2-4st)\xi_0]
    [tp_{{\rm F}-}-p_{{\rm F}-}^2+4s^2+2st]
        \over[t^2-p_{{\rm F}-}^2+4s^2+4st][tp_{{\rm F}+}+(p_{{\rm F}-}^2-4s^2)\xi_0]
    [tp_{{\rm F}-}+p_{{\rm F}-}^2-4s^2-2st]}\right\},
      \nonumber
\end{eqnarray}
\begin{eqnarray}
  f_1^{(b)}(t,s)=-{t\over2}~\!p_{{\rm F}-}(1+1/\xi_0^\prime)+2s^2+{1\over2}(t-2s)p_{{\rm F}+}
  -st+{t\over2\xi_0^\prime}p_{{\rm F}-}-sp_{{\rm F}-}\xi_0^\prime
  \phantom{aaaaaaaaaaaaaa}\nonumber\\
  +{t^2\over2}\ln\!\left(1+{p_{{\rm F}-}\over t}\right)
  +{p_{{\rm F}-}^2\over2}\ln\!\left({t-\xi_0^\prime p_{{\rm F}-}\over t+p_{{\rm F}-}}\right)
  -{t^2\over2}\ln\!\left(1+{p_{{\rm F}+}-2s\over t}\right)
  \phantom{aaaaaaaaaaaaa}~\!\nonumber\\
    +{1\over4}(p_{{\rm F}+}^2-4s^2+4st)
    \left\{-2\ln\!\left({2s+p_{{\rm F}+}\over4s\xi_0^\prime-p_{{\rm F}-}}\right)
    \right.\phantom{aaaaaaaaaaaaaaaaaaaaaaaaaaaaaa}~\!\nonumber\\
    \left.+\ln{[p_{{\rm F}+}^2-4s^2+4st-t^2][tp_{{\rm F}+}+p_{{\rm F}+}^2-4s^2+2st]
      [tp_{{\rm F}-}+(p_{{\rm F}+}^2-4s^2)\xi_0^\prime]\over
      [t^2-(p_{{\rm F}+}^2-4s^2+4st)\xi_0^{\prime2}][tp_{{\rm F}+}-p_{{\rm F}+}^2+4s^2-2st]
      [(p_{{\rm F}+}^2-4s^2+4st)\xi_0^\prime-tp_{{\rm F}-}]}\right\},\nonumber
\end{eqnarray}
\begin{eqnarray}
  f_1^{(c)}(t,s)=-tp_{{\rm F}-}+{1\over2}(p_{{\rm F}-}^2-t^2)
  \ln\!\left({t-p_{{\rm F}-}\over t+p_{{\rm F}-}}\right).
  \phantom{aaaaaaaaaaaaaaaaaaaaaaaaaaaaaaaaa}\nonumber
\end{eqnarray}
and
\begin{eqnarray}
  f_2^{(a)}(t,s)={t\over2}p_{{\rm F}-}+2s^2+{1\over2}(2s-t)p_{{\rm F}+}-st
  -\xi_0^\prime sp_{{\rm F}-}\phantom{aaaaaaaaaaaaaaaaaaaaaaaaaaaaaaaaaa}\nonumber\\
  +{1\over2}p^2_{{\rm F}-}\ln\!\left({t-\xi_0^\prime p_{{\rm F}-}\over t-p_{{\rm F}-}}\right)
  -{1\over2}t^2\ln\!\left({t-2s-p_{{\rm F}+}\over t-p_{{\rm F}-}}\right)
  \phantom{aaaaaaaaaaaaaaaaaaaaaaaaaaaa}\nonumber\\
  +{1\over4}(p_{{\rm F}+}^2-4s^2+4st)\left\{
  2\ln\!\left({2s+p_{{\rm F}+}\over p_{{\rm F}-}}\right)\right.
  \phantom{aaaaaaaaaaaaaaaaaaaaaaaaaaaaaaaaaaa}\nonumber\\
  \left.+\ln{[t^2-p_{{\rm F}+}^2+4s^2-4 s t][tp_{{\rm F}+}-p_{{\rm F}+}^2+4s^2-2st]
    [tp_{{\rm F}-}+(p_{{\rm F}+}^2-4s^2)\xi_0^\prime]\over
    [t^2-(p_{{\rm F}+}^2-4s^2+4st)\xi_0^{\prime2}][tp_{{\rm F}+}+p_{{\rm F}+}^2-4s^2+2st]
    [tp_{{\rm F}-}-(p_{{\rm F}+}^2-4s^2+4st)\xi_0^\prime]}\right\},\nonumber
\end{eqnarray}
\begin{eqnarray}
  f_2^{(b)}(t,s)=-2s^2-ts+sp_{{\rm F}-} + {t\over2}p_{{\rm F}-}
  -{t\over2\xi_0}p_{{\rm F}+} -\xi_0sp_{{\rm F}+}
  \phantom{aaaaaaaaaaaaaaaaaaaaaaaaaaaaaa}\nonumber\\
  -{t^2\over2\xi_0^2}\ln\!\left(1-{\xi_0p_{{\rm F}+}\over t}\right)
  +{t^2\over2}\ln\!\left(1+{2s -p_{{\rm F}-}\over t}\right)
  \phantom{aaaaaaaaaaaaaaaaaaaaaaaaaaaaa}~\nonumber\\
  +{1\over4}(p_{{\rm F}-}^2-4s^2-4st)\left\{
  -2\ln\!\left({2s-p_{{\rm F}-}\over p_{{\rm F}+}}\right)\right.
  \phantom{aaaaaaaaaaaaaaaaaaaaaaaaaaaaaaaa}~\nonumber\\
  \left.+\ln{[t^2-(p_{{\rm F}-}^2-4s^2-4st)\xi_0^2]
    [tp_{{\rm F}+}-(p_{{\rm F}-}^2-4s^2-4st)\xi_0][tp_{{\rm F}-}+p_{{\rm F}-}^2-4s^2-2st]
    \over[t^2-p_{{\rm F}-}^2+4s^2+4st][tp_{{\rm F}+}+(p_{{\rm F}-}^2-4s^2)\xi_0]
         [tp_{{\rm F}-}-p_{{\rm F}-}^2+4s^2+2st]}\right\}\nonumber\\
  -{t\over2}p_{{\rm F}+}(1-1/\xi_0)-{1\over2}
  p_{{\rm F}+}^2\ln\!\left({\xi_0p_{{\rm F}+}-t\over p_{{\rm F}+}-t}\right)
  -{t^2\over2}\ln\!\left(1-{p_{{\rm F}+}\over t}\right) + 
  {t^2\over2\xi_0^2}\ln\!\left(1-{\xi_0p_{{\rm F}+}\over t}\right),
  \phantom{a}\nonumber
\end{eqnarray}
\begin{eqnarray}
  f_2^{(c)}(t,s)=-tp_{{\rm F}+} + {1\over2}(p_{{\rm F}+}^2-t^2)
  \ln\!\left({t-p_{{\rm F}+}\over t+p_{{\rm F}+}}\right).
  \phantom{aaaaaaaaaaaaaaqaaaaaaaaaaaaaaaaaa}\nonumber
\end{eqnarray}
In these formulae
\begin{eqnarray}
  \xi_0\equiv{p^2_{{\rm F}-}-p_{{\rm F}+}^2-4s^2\over4s p_{{\rm F}+}}~\!,
  \phantom{aaaa} \xi^\prime_0\equiv
          {p^2_{{\rm F}-}-p_{{\rm F}+}^2+4s^2\over4s p_{{\rm F}-}}~\!.
          \label{eqn:xi0xi0prime}
\end{eqnarray}
Once the functions $f_1(t,s)$ and $f_2(t,s)$ are given in their analytic
forms, the functions $K^{(1)}(p_{{\rm F}-},p_{{\rm F}+})$ and
$K^{(2)}(p_{{\rm F}-},p_{{\rm F}+})$ can be evaluated using
the Mathematica package built-in instruction for numerical integration
over a specified domain.

Evaluation of the contribution to the energy density  of the
interactions proportional to the couplings $C_2$ and $C_2^\prime$ is
straightforward (no complicated integrals are involved). The result is
\begin{eqnarray}
 && {E_\Omega^{(C_2)}\over V}={C_2\over240\pi^4}~\!p_{{\rm F}-}^3p_{{\rm F}+}^3
  (p_{{\rm F}-}^2+p_{{\rm F}+}^2)~\!,\nonumber\\
   && {E_\Omega^{(C_2^\prime)}\over V}={C_2^\prime\over120\pi^4}\left[
    p_{{\rm F}+}^8+p_{{\rm F}-}^8
    +{1\over2}~\!p_{{\rm F}+}^3p_{{\rm F}-}^3(p_{{\rm F}+}^2+p_{{\rm F}-}^2)\right].
  \phantom{aaa}
  \label{eqn:C2C2PrimePolContrs}
\end{eqnarray}
These formulae agree for $p_{{\rm F}-}=p_{{\rm F}+}=k_{\rm F}$ with the ones
for $g_s=2$ obtained in \cite{HamFur00}.

\begin{figure}
\psfig{figure=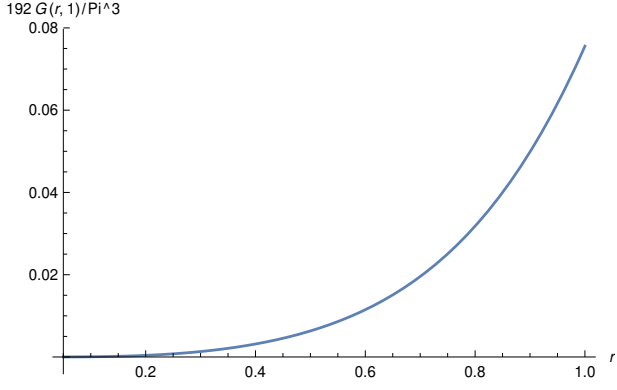,width=8.cm,height=5.0cm} 
\caption{Plot of the function $(192/\pi^3)G(r,1)$.}
\label{fig:Gplot}
\end{figure}

\begin{figure}
\psfig{figure=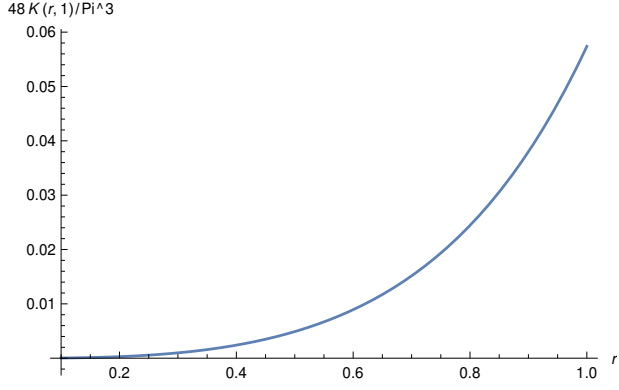,width=8.cm,height=5.0cm} 
\caption{Plot of the function $(48/\pi^3)K(r,1)$.}
\label{fig:Kplot}
\end{figure}

Combining (\ref{eqn:C03ppEnergyPol}) with (\ref{eqn:C03phEnergyPol}) and
(\ref{eqn:C2C2PrimePolContrs}), adding
the result (\ref{eqn:OthOrder}) (for $\alpha=+,-$), the known order
$k_{\rm F}^5(k_{\rm F}R)$ contribution $C_0p_{{\rm F}-}^2p_{{\rm F}+}^2/36\pi^4$,
the contribution of order $k_{\rm F}^5(k_{\rm F}R)^2$
\begin{eqnarray}
{64 m_fC_0^2\over(2\pi)^6\hbar^2}
  \int_0^{s_{\rm max}}\!ds~\!s^2~\!{1\over4\pi}\int\!d^3\mathbf{t}~\!
  \theta~\!\theta\left(-\Lambda+g_{\rm fin}(t,s)+{t^2\over\Lambda}\right),
  \label{eqn:SecondOrder}
  \end{eqnarray}
obtained in \cite{CHWO1} and finally expressing the couplings
$C_0$, $C_2$ and $C_2^\prime$ in terms of the $s$ and $p$-wave scattering
lengths $a_0$, $a_1$ and the $s$-wave effective radius $r_0$
using (\ref{eqn:CouplingsDetermined}) one easily finds
(using the results of \cite{CHWO1}) that up to the order
$k_{\rm F}^5(k_{\rm F}R)^3$ all terms diverging with $\Lambda\rightarrow\infty$
cancel out. One observes that the finite 
contribution arising in (\ref{eqn:SecondOrder}) from the term proportional to
$1/\Lambda$ after it is multiplied by the term $\propto\Lambda$ present
in $C_0^2$ cancels against the finite term $-2t^2$ arising from
squaring the function (\ref{eqn:g(t,s)General}) 
in the contribution of the  ``particle-particle'' diagram. Such terms
must cancel because they would be absent had one used Dimensional Regularization
instead of the cutoff $\Lambda$.
Defining then 
\begin{eqnarray}
  G(p_{{\rm F}-},~\!p_{{\rm F}+})=G^{(1)}_{\rm fin}(p_{{\rm F}-},~\!p_{{\rm F}+})
  +G^{(2)}(p_{{\rm F}-},~\!p_{{\rm F}+})~\!,\nonumber\\
  K(p_{{\rm F}-},~\!p_{{\rm F}+})=K^{(1)}(p_{{\rm F}-},~\!p_{{\rm F}+})
  +K^{(2)}(p_{{\rm F}-},~\!p_{{\rm F}+})~\!,
\end{eqnarray}
where $G^{(1)}_{\rm fin}(p_{{\rm F}-},~\!p_{{\rm F}+})$ is given by the formula
(\ref{eqn:G1FunctionDef}) with $g(t,s)$ replaced by $g_{\rm fin}(t,s)$
one arrives at the final formula
\begin{eqnarray}
  {E_\Omega\over V}={1\over6\pi^2}~\!{\hbar^2\over2m_f}\left\{
    {3\over5}\left(p_{{\rm F}-}^5+p_{{\rm F}+}^5\right)
    +{4\over3\pi}~\!p_{{\rm F}-}^3p_{{\rm F}+}^3~\!a_0+{192\over\pi^2}~\!a_0^2~\!
    J(p_{{\rm F}-},~\!p_{{\rm F}+})\right.\phantom{aaaaaaaa}~\!\nonumber\\
    +{384\over\pi^3}~\!a_0^3~\!G(p_{{\rm F}-},~\!p_{{\rm F}+})
    -{96\over\pi^3}~\!a_0^3~K(p_{{\rm F}-},~\!p_{{\rm F}+})\phantom{aaaa}~\!\nonumber\\
    +{1\over10\pi}~\!a^2_0r_0~\!p^3_{{\rm F}-}p^3_{{\rm F}+}\left(
    p_{{\rm F}-}^2+p_{{\rm F}+}^2\right)\phantom{aaaaaaaaaaaaa}~
    \label{eqn:Spin1/2Complete}\\
    \left.+
    {1\over5\pi}~\!a_1^3\left[2p^8_{{\rm F}-}+2p^8_{{\rm F}+}
      +p^3_{{\rm F}-}p^3_{{\rm F}+}\left(
      p_{{\rm F}-}^2+p_{{\rm F}+}^2\right)\right]\right\}.\nonumber
\end{eqnarray}
The function $J(p_{{\rm F}-},p_{{\rm F}+})$ is defined in \cite{CHWO1}.
It is also easy too see that  $J(p_{{\rm F}-},~\!p_{{\rm F}+})=p_{{\rm F}+}^7J(r,1)$,
$G(p_{{\rm F}-},~\!p_{{\rm F}+})=p_{{\rm F}+}^8G(r,1)$ and
$K(p_{{\rm F}-},~\!p_{{\rm F}+})=p_{{\rm F}+}^8K(r,1)$.
The plot of the function $J(r,1)$
has been given in \cite{CHWO1}. The  functions $(192/\pi^3)G(r,1)$ and
$(48/\pi^3)K(r,1)$ are shown here in Figures \ref{fig:Gplot} and
\ref{fig:Kplot}, respectively.

In the limit $p_{{\rm F}-}=p_{{\rm F}+}=k_{\rm F}$ the result
(\ref{eqn:Spin1/2Complete}) should coincide with 
\begin{eqnarray}
  {E_\Omega\over V}={1\over6\pi^2}~\!{\hbar^2\over2m_f}\left\{
  g_s~\!{3\over5}~\!k_{\rm F}^5+g_s(g_s-1)~\!{2\over3\pi}~\!k_{\rm F}^6a_0
  +g_s(g_s-1)~\!{4(11-2\ln2)\over35\pi^2}~\!k_{\rm F}^7a_0^2\right.
  \phantom{aaaaa}\nonumber\\
  +[g_s(g_s-1)~\!N_1+g_s(g_s-1)(g_s-3)~\!N_2]~\!k_{\rm F}^8~\!a_0^3
  \phantom{aaaaaaaaaaa}\nonumber\\
  \left.+g_s(g_s-1)~\!{1\over10\pi}~\!k^8_{\rm F}~\!a_0^2r_0
  +g_s(g_s+1)~\!{1\over5\pi}~\!k^8_{\rm F}~\!a_1^3\right\},\phantom{aaaaaaaa}~
\end{eqnarray}
for $g_s=2$ given in \cite{HamFur00} and \cite{WeDrSch}
with $N_1=0.07550\pm0.00003$ and $N_2=0.05741\pm0.00002$ in \cite{HamFur00},
and $N_1=0.0755732$ and $N_2=0.0573879$ in \cite{WeDrSch}.
Numerical evaluation of the functions $(192/\pi^3)G(1,1)$ and
$(48/\pi^3)K(1,1)$ - the endpoints in Figures \ref{fig:Gplot} and \ref{fig:Kplot},
respectively -
gives $N_1=0.0755617$ and $N_2=0.057387$ in good agreement with the numbers
obtained in \cite{HamFur00} and \cite{WeDrSch}. (In \cite{CHWO1}
it has been found that $J(1,1)=0.0114449$ which with high accuracy equals
$(11-2\ln2)/840$).

Expressed in terms of $k_{\rm F}=(3\pi^2N/V)^{1/3}$ and $r$
the formula (\ref{eqn:Spin1/2Complete}) takes the form
\begin{eqnarray}
  {E_\Omega\over V}={k_{\rm F}^3\over3\pi^2}~\!{\hbar^2k_{\rm F}^2\over2m_f}~\!{3\over5}
  \left\{{1\over2}\left(1+r^5\right)\!\left({2\over1+r^3}\right)^{5/3}
  +{10\over9\pi}~\!r^3\!\left({2\over1+r^3}\right)^2\left(k_{\rm F}a_0\right)
  \right.\phantom{aaaaaaaaaaaaaaaaaa}\nonumber\\
  +{160\over\pi^2}\left({2\over1+r^3}\right)^{7/3}\!
  J(r,1)\left(k_{\rm F}a_0\right)^2
  \phantom{aaaaaaaaaaaaaaaaaaaaaaaaaaaaaaaa}\label{eqn:Final}\\
  +{80\over\pi^3}\left({2\over1+r^3}\right)^{8/3}\!
  \left[4~\!G(r,1)-K(r,1)\right]\left(k_{\rm F}a_0\right)^3
  \phantom{aaaaaaaaaaaaaaaaaaaaaa} \nonumber\\
  \left.+{1\over12\pi}\left({2\over1+r^3}\right)^{8/3}\!\!
  \left[r^3(1+r^2)~\!(k^3_{\rm F}a_0^2r_0)
    +2(2+2r^8+r^3+r^5)(k_{\rm F}a_1)^3\right]\right\}.\nonumber
\end{eqnarray}
The third order corrections computed in this work (the last two lines in the
above formula) are rather small. For $r_0=a_1=0$ (i.e. without the contribution
on the dimension $R^{-6}$ operators) the ratio of the order $(k_{\rm F}a_0)^3$
contribution to the first term in the curly brackets
increases from 0.00003 at $k_{\rm F}a_0=0.1$ to $0.03$ at $k_{\rm F}a_0=1$.
This can be compared to the analogous ratio of the order $(k_{\rm F}a_0)^2$
term which at $r=1$ increases from 0.00185 to 0.185.
These ratios decrease further with decreasing $r$ (increasing polarization)
and become exactly zero at $r=0$ due to the Pauli exclusion which
forbids any contribution to the ground state energy to be generated
by the interaction operator proportional to $C_0$.

\begin{figure}
\psfig{figure=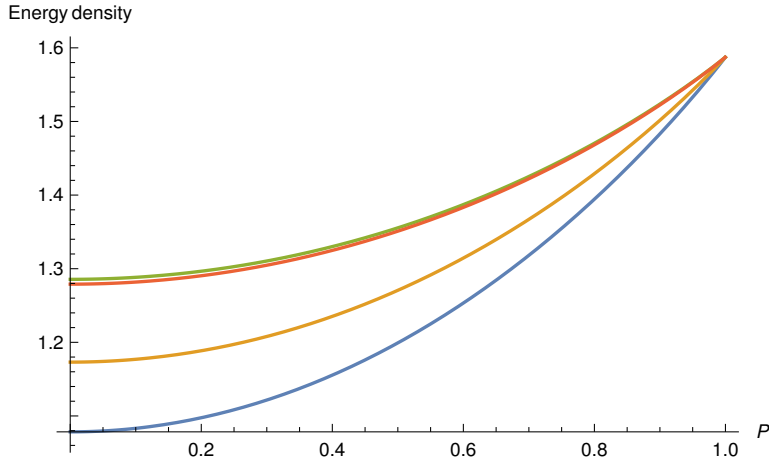,width=10.cm,height=6.0cm} 
\caption{Energy density $E_\Omega/V$ in units $(3/5)\hbar^2k^5_{\rm F}/6\pi^2m_f
  =(N/V)(\hbar^2k^2_{\rm F}/2m_f)(3/5)$
  of the gas of spin $1/2$ fermions as a function of
  its polarization $P=(N_+-N_-)/N$  for different values
  (from below) 0.2 (blue), 0.4 (yellow), 0.6 (green) of the expansion
  parameter $k_{\rm F}a_0$. The last curve (red) corresponding to $k_{\rm F}a_0=0.6$
shows the same quantity but without the order $(k_{\rm F}a_0)^3$ correction.}
\label{fig:Energy1}
\end{figure}

The plot of the system's ground state energy density
as a function
of the polarization $P$ related to $r$ by $r(P)=((1-P)/(1+P))^{1/3}$
is shown in Fig. \ref{fig:Energy1} for three
different values of the expansion parameter $k_{\rm F}a_0$ (keeping
$r_0=a_1=0$). All curves merge at $P=1$ as a result of the
Pauli exclusion principle.
The curve corresponding to $k_{\rm F}a_0=0.6$ can be directly
compared to the lowest curve shown  in Fig. 3 of ref. \cite{QMC10}
which shows a numerical estimate of the exact ground
state energy obtained using the Quantum Monte Carlo method
for a specific model repulsive potential. Consistently with the
comparison of the ground state energies of the unpolarized system
($P=0$ or $r=1$) made in Fig. 2 of ref. \cite{QMC10},
our green curve (for $k_{\rm F}a_0=0.6$) is systematically below its
counterpart in Fig. 3 of ref. \cite{QMC10} but the comparison
with the red curve of Figure \ref{fig:Energy1} shows that the
third order correction computed in this work has the tendency to reduce
the difference between the perturbative and Monte Carlo estimates.
In general, the comparison with the results of ref. \cite{QMC10} show
that the perturbative expansion is reliable up to $k_{\rm F}a_0\simlt0.5$.
\vskip0.1cm

\noindent{\bf Summary.}
In this work we have reproduced the third order formula for the ground-state
energy of the unpolarized gas of spin $s$ fermions and extended it to the
case of the arbitrarily polarized gas of spin $1/2$ fermions. We have checked
the cancellation of all ultraviolet divergences occurring when the result is
expressed in terms of the $s$-wave scattering length $a_0$ and worked out
analytically the most important integrals occurring in the computation of the
relevant Feynman (Hughenholtz) diagrams. This allowed to compute the remaining
integrals numerically using the standard Mathematica built-in routines; the
resulting final third order formula for the ground state energy of the
arbitrarily polarized gas of spin $1/2$ fermions is given in terms of two new
functions of the system's polarization for which the convenient interpolating
formula can be easily obtained. The numerical results suggest that for
$k_{\rm F}R\simlt0.5$ the perturbation series for the ground-state energy
is well convergent but is not reliable for higher values of the expansion
parameter for which the system is expected to exhibit the phase transition
to the ordered phase. One may hope, however, that supplemented
with a reliable extrapolation procedure the perturbative series will be able
to give valuable information about the nature of the phase transition.
Further extension of our work to the
case of arbitrary mixture of different spin projections of spin $s$
fermions (in the spirit of \cite{PECABO}) is straightforward.

\end{document}